\documentclass[sigconf,nonacm]{acmart}
\usepackage{amsmath}
\usepackage{booktabs}
\usepackage{tabularx}
\usepackage{tikz}
\usetikzlibrary{arrows.meta,positioning,fit,backgrounds}
\setcopyright{none}
\acmYear{2026}
\acmDOI{}
\acmISBN{}
\newcommand{\unknown}{\mathsf{unknown}}

\begin{document}
\title[Source Information in Android Agents]{From Observations to Authorized Actions: Measuring Source Information in Android Agents}

\author{Hyobin Ahn}
\affiliation{%
  \institution{Kangwon National University}
  \city{Chuncheon}
  \country{Republic of Korea}}
\email{gyqlsdks@kangwon.ac.kr}
\author{Dongjae Lee}
\affiliation{%
  \institution{Kangwon National University}
  \city{Chuncheon}
  \country{Republic of Korea}}
\email{dongjae.lee@kangwon.ac.kr}
\renewcommand{\shortauthors}{Ahn and Lee}

\begin{abstract}
Android agents read content from applications and web pages and act on it with the user's authority, yet they cannot reliably tell where that content came from---its \emph{source}, the application or document that supplied it. The WebView accessibility fields we inspected expose the host application but omit the document origins available inside the browser, so an agent reading through this interface cannot check which document supplied an object; a limited reproduction on an unmodified WebView yields the same finding. We build and separately evaluate two prototypes to examine how source information supports authorization. Our application-source monitor withholds side-effecting calls whose signed source record contains an untrusted application. With the source set derived from the OS accessibility log, it blocked every selected attack write in the tested runs, but it also withheld benign sends issued after reading untrusted content, flagged benign tasks that use external content, and carried source records into a later clean task. Our WebView source-reference gate uses records from a modified browser to check object origins and reference lifetime through direct calls; in a separate 24-trial evaluation, the live agent selected no disallowed object. Supporting benchmark and planner measurements characterize textual argument matches and attacker-target selection, without establishing causal dataflow or a controlled model ranking. The prototypes are not an integrated defense: the WebView gate still lacks authenticated browser delivery and atomic capture of selected bytes and origin. Together, the results show which source information the tested policies need and why it must stay linked to the object used in execution.
\end{abstract}

\maketitle

\section{Introduction}
An Android agent reads web pages, notifications, and application screens, then invokes functions with the user's authority. A message recipient or body may come from this observed content rather than from the user's request. In our prototypes, the agent invokes functions in another application through Android AppFunctions~\cite{appfunctions}. Caller authentication identifies the requesting application but does not establish the source of the content in the request. A message copied from an embedded advertisement and one supplied directly by the user can reach the same execution boundary under the same caller identity.

\begin{figure*}[t]
\centering
\begin{tikzpicture}[font=\small,
 box/.style={draw,rounded corners=2pt,align=center,inner sep=5pt,text width=0.18\textwidth},
 arr/.style={-{Latex[length=1.8mm]},thick}]
\node[box] (content) at (0,0) {Application content\\including WebView};
\node[box] (observe) at (3.95,0) {OS accessibility\\delivery};
\node[box] (agent) at (7.9,0) {Agent\\AppFunctions caller};
\node[box] (broker) at (7.9,-2) {Platform broker\\request checks};
\node[box] (provider) at (3.95,-2) {ChatApp provider\\in-memory messages};
\draw[arr] (content)--(observe);
\draw[arr] (observe)--(agent);
\draw[arr] (agent)--node[right]{call}(broker);
\draw[arr] (broker)--(provider);
\node[draw,align=left,text width=0.88\textwidth,inner sep=5pt] at (3.95,-4.1)
{\textbf{Application-source case:} our OS instrumentation records source packages at accessibility delivery; the broker and provider check recent history.\\[3pt]
\textbf{WebView case:} our modified browser supplies object/origin records to the OS reference table; the broker and provider check a selected-object reference. Browser authentication and atomic capture remain unimplemented.};
\end{tikzpicture}
\caption{Observation and execution path used by our two separate prototypes. The source records and checks differ between cases.}
\label{fig:overview}
\end{figure*}
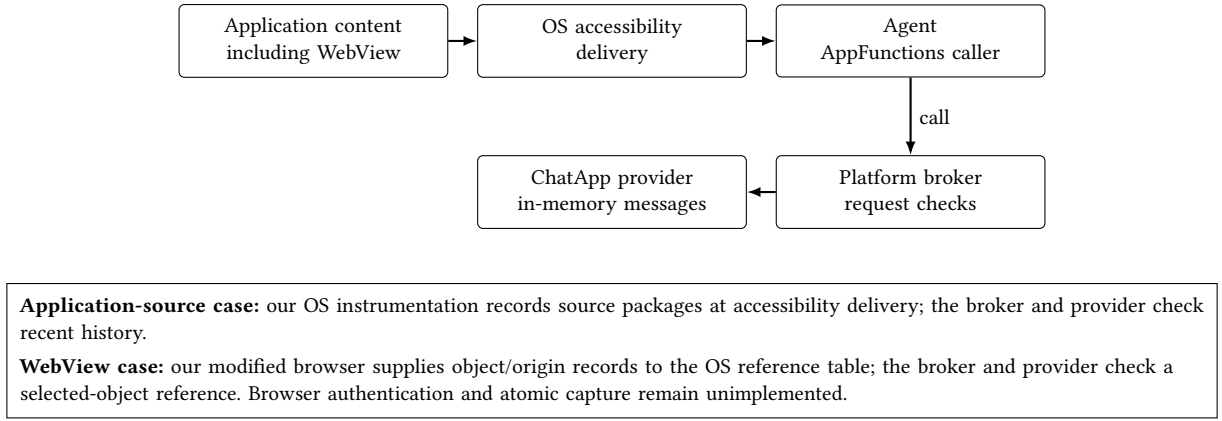

\paragraph{System model.} The agent observes application content through Android accessibility, including web content rendered in WebView; Android GUI agents commonly read the accessibility tree, often together with screenshots~\cite{androidworld,appagent,autodroid}. It acts through AppFunctions: the agent is the \emph{caller}, the platform's \emph{broker} mediates the request, and the \emph{provider} is the application that executes the function. Our experimental provider, ChatApp, stores messages in memory. We add source recording and execution checks to this path; Figure~\ref{fig:overview} shows where the two prototypes obtain their source information.

Indirect prompt injection shows why observed content cannot automatically be treated as instructions~\cite{greshake}. Consider the policy ``share only content selected from an approved source.'' Enforcing it requires evidence beyond the planner's claim: which source supplied the selected object, and whether the action uses that same object. Because a trusted host application can display untrusted content, trust in the host does not determine eligibility. We therefore distinguish whether a source is eligible from whether the user authorized the action, since an approved source does not authorize every recipient or operation.

We organize the study around two questions. What can recent application history tell an execution check? What additional information is needed to check the origin of a selected web object? We implement an application-source monitor and a WebView source-reference gate to examine these questions separately. The latter uses a modified browser and simulated approval; its implementation status is specified in \S\ref{sec:design}.

Consider a host application displaying two embedded documents. A sharing policy permits a selected object from origin $P$ but not an otherwise identical object from $Q$. The host package and selected bytes alone do not express that distinction. This illustrative case motivates our selected-object tests. The application-source monitor asks a different question---which applications the agent recently observed---so it is evaluated under a separate threat model and workload.

\paragraph{Contributions.} We make three contributions:
\begin{enumerate}
\item We measure an origin-attribution gap in the inspected WebView accessibility metadata, compare it with browser-side origin information, and reproduce the finding on a limited unmodified configuration (\S\ref{sec:measurement}).
\item We implement and separately evaluate two authorization prototypes (\S\ref{sec:nativeeval}, \S\ref{sec:webvieweval}). One shows how application history can restrict benign tasks. The other tests origin checks on selected objects, but still lacks authenticated delivery and atomic capture of the object--origin binding.
\item We provide supporting measurements of argument matching and planner behavior that motivate checking actions independently of model-selected instructions (\S\ref{sec:workload}, \S\ref{sec:susc}). Argument matches are textual rather than causal, and the planner trials are exploratory rather than a controlled model ranking.
\end{enumerate}

The study concerns observation interfaces and execution checks. We do not pool the prototypes' outcomes into one defense rate. Table~\ref{tab:roadmap} maps the measurements to the two prototypes. The WebView measurement motivates the selected-object case. The benchmark and planner measurements motivate checking individual actions independently of the planner. We present the application-level case first, then the selected-object case. Together, they illustrate why a check must identify the application or document required by its policy and link that identity to the object it executes on.

\section{Background and Threat Model}
\subsection{Source identities and related properties}
\label{sec:scope}
We use \emph{provenance} for information about where observed content came from. A \emph{source identity} identifies an application or document; it need not establish how an argument was derived. Below, source identity means an application package in the native case and a document origin in the WebView case. We also distinguish three related properties: argument derivation, user authorization, and freshness at execution. The distinction determines what an execution check can infer from an observation.

\begin{itemize}
\item \textbf{Native application provenance}: which installed application supplied an observed node (the Android package behind an accessibility node). Observable at the OS accessibility-delivery point; it does \emph{not} identify a document origin within a browser.
\item \textbf{Web document origin}: the browser-computed security origin (scheme, host, port, or an opaque/inherited origin) of a rendered document, including an embedded cross-origin frame. Available inside the browser; the question is whether it is preserved when content is delivered to an external consumer.
\item \textbf{Argument derivation}: whether a value placed in a side-effecting tool argument came from the user request or from environment content. A workload property, independent of whether that content is malicious.
\item \textbf{User authorization}: whether the user approved \emph{this} action's scope (object, recipient, operation). Orthogonal to source: approved-source content may still be sent to an unapproved recipient.
\item \textbf{Freshness}: whether the object and bytes bound at approval are still the ones used at execution, across navigation or in-place change (a time-of-check/time-of-use property).
\end{itemize}

Figure~\ref{fig:overview} shows the shared execution path and the different source records used in each experiment. Table~\ref{tab:roadmap} summarizes the questions and evaluation settings. The detailed evidence map remains in Appendix~\ref{app:tables}.

\begin{table*}[t]
\caption{Reading map. Table~\ref{tab:evmap} in Appendix~\ref{app:tables} gives the detailed evidence map.}
\label{tab:roadmap}
\centering\small
\begin{tabularx}{\textwidth}{p{0.19\textwidth}XXl}
\toprule
Part & Question & Setting and evidence & Section \\
\midrule
WebView measurement & Does accessibility preserve document origin? & Two-origin fixture; browser ground truth and accessibility collectors & \ref{sec:measurement} \\
Argument matching & Which argument values match environment text but not the prompt? & AgentDojo text matching, not causal derivation & \ref{sec:workload} \\
Planner behavior & Does the planner select the attacker's target? & Exploratory injection trials; selection measured separately from writes & \ref{sec:susc} \\
Application-source monitor & Can recent application identities constrain writes? & Attacker controls source-app content; OS package records trusted; native fixture with Qwen2.5-7B & \ref{sec:nativeeval} \\
WebView source-reference gate & Can execution check the selected object's origin and lifetime? & Attacker controls embedded document; cooperating browser and simulated approval; direct calls and limited live trials & \ref{sec:webvieweval} \\
\bottomrule
\end{tabularx}
\end{table*}

\subsection{Origins, frames, and accessibility}
A tuple origin comprises a scheme, host, and port. A document's security origin is not necessarily obtained by parsing its displayed URL: inherited origins and opaque origins require the browser's origin semantics~\cite{html}. A sandboxed document without the \texttt{allow-same-origin} flag can have an opaque origin. Consequently, adding an arbitrary URL string to an accessibility node is not a complete origin signal.

Android accessibility nodes expose text, package and class information, tree relations, extras, and a list of keys for requesting additional data~\cite{a11y}. The package identifies the application supplying the node; it is not a statement that every displayed sentence was authored by that application. Chromium's accessibility implementation also uses extras for geometry and semantic roles~\cite{chromium}. Frame identification and origin identification are separate: an iframe role can support conservative treatment of all embedded content without identifying which security origin owns that content.

\subsection{Source eligibility and action authorization}
\label{sec:policy}
An object's source and the user's authority to act on it answer different questions. Even perfect attribution to an approved origin does not authorize an arbitrary transfer or message. We separate two predicates:
\[
 \mathsf{Allow}(a)=\mathsf{SourceEligible}(a)\land\mathsf{TaskAuthorized}(a).
\]
The first checks source requirements for data used in action $a$ (for origins, membership in an approved-origin set); the second checks a user-issued action scope. Either may fail independently. Selected content may come from an approved site but be proposed for an unapproved recipient. Conversely, the recipient may be authorized while a substituted attachment comes from a disallowed document.

Origin metadata is therefore evidence about an object, not permission for every action involving it. Appendix~\ref{app:proto} gives the restricted-input argument explaining why package and metadata-key identity alone cannot implement an origin-sensitive policy.
\subsection{Attacker and trusted components}
The motivating attacker controls the body of an embedded document that a legitimate page displays. The attacker may insert instructions or values into that document, but does not modify the user's request, the host application, the browser implementation, the OS, or an executing provider. We model this separation using different local HTTP ports in the test fixture, so the embedded document and host have distinct origins. It does not establish how an attacker would obtain an embedding position in a particular production site.

\textbf{WebView case.} For the WebView design, we assume a trusted browser, OS delivery path, selection interface, and cooperating action provider. These are design assumptions, not properties established by our prototype. The planner may propose arbitrary calls and arguments, including calls induced by hostile content. The provider must therefore check authority independently of the planner. We exclude browser compromise, hostile accessibility providers impersonating a trusted browser, and a malicious action provider from this threat model.

\textbf{Application-source case.} For the native-source variant, we let the attacker control content in a source \emph{application} whose nodes the agent reads through accessibility. We trust the OS to record (stamp) the originating package at delivery and exclude platform compromise and forged OS stamps from this threat model. The application-source monitor and WebView gate are evaluated under these separate trust assumptions (\S\ref{sec:nativeeval}, \S\ref{sec:webvieweval}).

\textbf{Common setup.} In both prototypes, the agent calls a provider through Android AppFunctions~\cite{appfunctions}. The platform authenticates the calling application. We modify the broker and provider on a patched platform to add source recording and enforcement; this is not the shipping framework. We evaluate calls to the experimental ChatApp provider. AppFunctions, Chromium, AgentDojo, and the tested language models are existing components. The corresponding sections describe our additions and measurement procedures.

An authorized accessibility consumer is expected to read rendered content, including cross-origin frames. We ask whether the delivered representation also retains the metadata needed to distinguish document origins. This question does not equate a trusted origin with trustworthy instructions.

\section{Related Work}
We relate the study to defenses against prompt injection, authorization at tool boundaries, browser origin policies, and Android provenance tracking. These areas address different parts of the path from observing content to executing an action.
\subsection{Indirect prompt injection and model-side defenses}
Indirect prompt injection places attacker instructions in data consumed by LLM-integrated applications~\cite{greshake}. Our focus is the platform metadata available to distinguish the \emph{sources} of that data. Benchmarks of indirect prompt injection in tool-integrated agents, such as InjecAgent~\cite{injecagent}, evaluate how often agents follow injected instructions. Our planner-susceptibility measurement (\S\ref{sec:susc}) uses a separate in-house fixture; it is not a run of InjecAgent.

Much defensive work acts on the model or the data pipeline. CaMeL separates privileged planning from untrusted data processing and enforces capability and information-flow policies in an interpreter~\cite{camel}. Fides enforces information-flow labels over agent data~\cite{fides}. AttriGuard attributes tool calls counterfactually with shadow replay and observation attenuation~\cite{attriguard}. We examine how source information can support checks independent of the planner. We do not evaluate comparative effectiveness against these systems.

Our experiments examine the acquisition of a source label at an Android observation interface, its binding to the object an action carries, and the lifetime of that binding across a native execution boundary. These questions concern the input to an enforcement mechanism.
\subsection{Authorization and confused-deputy failures at the tool boundary}
AgentSpec expresses runtime constraints through triggers, predicates, and actions~\cite{agentspec}. Intent-Governed Tool Authorization issues intent certificates for constrained execution~\cite{igac}. ``Capability Gates Are Not Authorization'' shows that frameworks may control tool availability without authorizing a specific call's arguments~\cite{scopegate}.

Adding a gate, certificate, or exact-payload check is thus not itself novel. We focus on obtaining source information and binding it to an action across a native interface. We evaluate a monitor using signed application-identity records, derived in one configuration from the OS accessibility log (\S\ref{sec:nativeeval}). Separately, our WebView prototype adds an origin check to otherwise identical capability checks (\S\ref{sec:onoff}). It assumes a cooperating browser; authenticated delivery remains a requirement.

Consent Integrity binds a user's approval to the canonical action and refuses swapped actions~\cite{consent}. Atomicity for Agents exposes time-of-check/time-of-use gaps between observation and action and detects changes after observation~\cite{atomicity}. A digest re-check detects bytes that differ from those approved. An immutable snapshot instead preserves the approved bytes without re-reading the object. We examine how to associate those bytes with an object and its origin across a native boundary that omits the needed source label.

\subsection{Browser-agent origin boundaries}
Two concurrent efforts constrain browser agents by origin and are the nearest prior work. Chrome's Agent Origin Sets restricts, inside the browser, which origins an agent task may read and act on~\cite{aos}. WebMCP-Phalanx binds each page-registered tool to its registering principal with cryptographic capability credentials. It carries provenance labels through the tool lifecycle and screens page content with a quarantined agent~\cite{phalanx}.

Both operate inside the browser, where origin identities are already available. Earlier WebView security work showed that the trust boundary between web content and the host application weakens inside WebView~\cite{webviewattacks}. We study native accessibility consumers receiving rendered WebView content. The fields we inspected omit per-node document origins (\S\ref{sec:measurement}). We also test how to bind a selected object's origin to execution by an external provider. We do not claim a stronger in-browser boundary.

\subsection{Android platform provenance and the accessibility surface}
Our application-source monitor builds on OS-level provenance tracking. TaintDroid instruments the Android runtime to track flows from sensitive sources to sinks for real-time privacy monitoring~\cite{taintdroid}. We record only the originating application package at accessibility delivery. This supports application-level authorization checks, not byte-level taint tracking or causal dataflow analysis.

Accessibility is a powerful and historically abused interface. Jang et al.\ showed that accessibility interfaces across operating systems can be abused to bypass security mechanisms~\cite{a11yattacks}. Cloak and Dagger shows that accessibility-service and overlay permissions together grant near-complete control of the Android UI feedback loop~\cite{cloakdagger}. This motivates a trust boundary around content received through accessibility. Source assertions at that boundary need authentication, rather than reliance on a model-supplied hint (an unverified source claim).

Action Rebinding examines the observation-to-action gap of Android GUI agents~\cite{rebinding}. Our lifetime tests examine document replacement with and without an origin change (\S\ref{sec:s3}, Table~\ref{tab:life}). We have not systematically distinguished rebinding failures from failures to enforce document-reference lifetime. We also inspect whether accessibility fields preserve document origin alongside text, roles, and other metadata (\S\ref{sec:measurement})~\cite{a11y,chromium,dumper}.

\section{Source Information at Observation Interfaces}
We begin with the source information available to an Android accessibility consumer: application identity is preserved across the interface, while document origin is absent from the WebView fields we inspect. We then report two supporting measurements of action arguments and planner behavior. They motivate execution checks but do not establish the origin gap or validate either prototype.

\subsection{Measuring the WebView boundary}
\label{sec:measurement}
The inspected accessibility fields expose the host application but omit the document origin available inside the browser. This matters when a policy must permit one embedded origin and reject another. We first establish this origin-attribution gap, then use a modified browser to supply the missing record in \S\ref{sec:webvieweval}.

\subsubsection{Environment and fixture}
We run the primary interface measurement on a modified AOSP emulator with security patch level June 5, 2026. The selected WebView provider is \path{com.android.webview}, version 145.0.7632.218, version code 763221807, targeting SDK 36. We use the AOSP WebView shell as the host. According to our experiment record, the research Chromium patch was not installed in this provider. In this modified environment, package metadata identifies the selected provider and version but does not verify byte-for-byte identity with a released binary.

The origin-attribution gap measurement and the origin-preserving prototype run in \emph{different} environments (Table~\ref{tab:env}); we report only the components actually run in each. We measure the interface directly, without a live agent, in both the primary run and the unmodified-platform reproduction; the prototype evaluation includes the limited live-agent runs of \S\ref{sec:live}.

\begin{table*}[t]
\caption{Environments for observation measurements and the WebView prototype.}
\label{tab:env}
\centering\small
\begin{tabularx}{\textwidth}{lXXX}
\toprule
Environment & Browser and platform & Collector & Provider / agent \\
\midrule
Primary gap measurement & \path{com.android.webview} 145.x; research browser patch absent; modified AOSP & Public-API service and framework callback & None \\
Unmodified reproduction & Stock \path{com.google.android.webview} 133.x; stock \texttt{google\_apis}, Android 16 & Public-API service in System WebView Shell & None \\
WebView prototype & Patched \texttt{content\_shell} 157.x; modified AOSP & System-server delivery and source-reference table & ChatApp in-memory store; minimal live agent \\
\bottomrule
\end{tabularx}
\end{table*}

To compare host and embedded content under distinct origins, we serve a host page at \url{http://127.0.0.1:8901/host.html} with an iframe loading \url{http://127.0.0.1:8902/evil.html}. We place a different marker in each document to identify its nodes in the collected records, and include a sentence requesting an action in the child. The sentence is fixture text, not evidence of an executed transfer. We serve both pages locally and expose them to the emulator with ADB reverse forwarding.

\subsubsection{Browser-side origin comparison}
\label{sec:interfaces}
We compared two extraction interfaces on the two-origin fixture. For browser-side ground truth, we ran a script in each frame. Each script read its own \texttt{location.origin} and reported the result by HTTP request to its own origin's server. The host reported \path{http://127.0.0.1:8901}; the embedded iframe reported \path{http://127.0.0.1:8902}. For these plain-HTTP frames, \texttt{location.origin} equals the document's security origin. Sandboxed or inherited-origin documents require the browser's internal security origin rather than a parsed URL origin. Those cases are outside this comparison.

\subsubsection{Collector and measurement limits}
To compare public-API observations with framework delivery, we collect the same fixture nodes through two paths. We first run a standalone accessibility service using only public interfaces (\texttt{getRootInActiveWindow}, depth-first \texttt{getChild}, \texttt{getExtras}, and \texttt{getAvailableExtraData}). We install it as an ordinary application, disable the framework instrumentation, and enable only this service for that run.

For the second path, we instrument the framework's accessibility-node delivery callback to enumerate the same fields on nodes handed to a consumer. The instrumentation covers single-result, list-result, and prefetch callbacks. We keep a separately gated text-substitution method disabled (\texttt{debug.pcaf.handles=0}; \texttt{pcaf} is the property namespace of our instrumentation) during collection; the instrumentation runs before that method, and no delivered node was modified.

Both collectors record the same geometry, role, and hint-text keys and the same advertised extra-data keys on the WebView content nodes. Neither exposes a document origin. The standalone run records one additional key, \texttt{cssDisplay}, which also carries no origin. Table~\ref{tab:counts} reports counts for the instrumented-callback trace.

We use the standalone service to check whether the observation depends on our framework instrumentation. These two collector runs share the research image and therefore do not themselves constitute an unmodified-platform reproduction; the separate reproduction in \S\ref{sec:stockrepro} addresses that limitation.

Neither collector establishes coverage of every delivery path or cached node. The logs lack stable node identifiers and traversal-completion records. Repeated callbacks therefore cannot be interpreted as unique-node coverage.

Our collector enumerates keys but truncates value representations to 80 characters and node text to 44 characters. It does not recursively serialize nested values. Our negative observation is limited to those keys and recorded representations. The supplied UIAutomator XML shows visual hierarchy, but serializes only selected fields. It cannot establish that unprinted extras are absent~\cite{dumper}.

\subsubsection{Observed accessibility metadata}
We recompute the descriptive counts in Table~\ref{tab:counts} from the raw log. We recorded 191 delivery records, including 164 for the shell package and 27 for other applications. A \texttt{chromeRole} key appears in 93 records. Each marker appears 13 times. These are repeated observations, not independent trials or distinct documents. We therefore do not attach confidence intervals to them.

For both marker nodes, the package is the shell, the class is \path{android.widget.TextView}, and the resource identifier is null. Their extras include six unclipped-geometry fields, a hint-text field, and a paragraph role. Their keys coincide, but their geometry values differ: for example, the recorded widths are 1063 and 379. The iframe container exposes \texttt{chromeRole=iframe}, and the child frame root exposes \texttt{rootWebArea}. These roles identify structure, not a document security origin.

Across the records containing a Web role, additional observed extras keys describe headings and supported HTML navigation elements. None of the enumerated keys is a document URL or security-origin key. Both marker nodes advertise text-character-location data. Across the whole trace, the advertised sets contain rendering information, character-location information, or no keys. We did not observe an advertised origin-retrieval key; we did not test arbitrary unadvertised requests.

All 11 recorded HTTP(S) URL occurrences are associated with the shell's address-bar resource identifier. This is host application UI, not per-document origin metadata for the child content. This count describes literal occurrences in the recorded output, not an exhaustive search.

\subsubsection{Unmodified-platform reproduction}
\label{sec:stockrepro}
To check whether the gap depends on the modified image, we repeated the measurement on an unmodified, Google-distributed stack. We used stock \path{com.google.android.webview} 133.0.6943.137 on a stock \texttt{google\_apis} image (Android~16). This provider is older than the primary one (133 vs.\ 145); we did not test a stock 145 provider. The System WebView Shell rendered the same two-origin fixture with its stock provider. We collected nodes with the same public-API service, the only non-stock component in this run.

The marker nodes again expose only geometry, role, \texttt{cssDisplay}, and hint-text keys. They include a generic \texttt{chromeRole} boundary, but no document-origin key: no enumerated extras key or advertised extra-data key names a document origin (origin-key occurrences: $0$), and the only URL-bearing value is the shell's address bar.
\begin{table}[t]
\caption{Document-origin information in the inspected fixture. Absence is scoped to the recorded fields and representations.}
\label{tab:gap_summary}
\centering\small
\begin{tabularx}{\linewidth}{Xl}
\toprule
Observation point & Origin observed? \\
\midrule
DOM: each plain-HTTP frame & Yes \\
Accessibility: public-API collector & No \\
Accessibility: framework collector & No \\
Accessibility: unmodified reproduction & No \\
\bottomrule
\end{tabularx}
\end{table}

\subsubsection{Interpretation}
Table~\ref{tab:gap_summary} summarizes the contrast: the plain-HTTP fixture exposes each origin inside the browser, but the inspected WebView nodes expose no document-origin field in the enumerated extras or advertised extra-data keys. The stronger statements that no accessibility mechanism can recover an origin, that the entire rendered tree was exhaustively captured, or that all WebView versions behave this way are not established. Nested, inherited, and sandboxed opaque origins need separate ground truth; parsing a URL would not suffice for those cases.

A consumer could use the iframe role to withhold all actions involving embedded content. The missing distinction matters when it wishes to permit one embedded origin and withhold another. A top-level URL cannot by itself make that distinction. This is an attribution prerequisite for a policy, not evidence that a particular consumer currently implements the policy incorrectly.

\subsection{Supporting measurement: argument matches}
\label{sec:workload}
This supporting measurement motivates the application-source case in \S\ref{sec:nativeeval}. Legitimate action arguments can contain values absent from the user's request. A rule that restricts external sources can therefore affect benign tasks. We examine how often sensitive arguments match environment text in AgentDojo~\cite{agentdojo}. We use all four task suites of version~v1 (workspace, Slack, travel, and banking). For each user task, we load its default environment and ground-truth tool calls. The counts below are our classification of that suite, not results reported by the benchmark.

We select ground-truth calls to functions with side effects, such as send, pay, transfer, schedule, create, post, add, delete, update, and share. We then extract sensitive arguments, such as recipient, amount, IBAN, account, URL, email, participants, phone, address, payee, and destination.

A value is a \emph{prompt match} if it appears in the user prompt. It is an \emph{environment-only match} if it appears only in the environment dump, and \emph{no match} if it appears in neither. We give prompt matches precedence when both contain the value and count list elements separately. These categories describe textual overlap, not the source an agent actually used.

To expose dependence on the definition of a sensitive argument, we report classifications under both a broader and a narrower field set. Under the broader field set, the classifier labels 44 of 81 argument instances (54.3\%) as environment-only matches. Under the narrower field set, 25 of 57 instances (43.9\%) match the environment dump but not the user prompt. We report these as two specification-dependent measurements, not as an interval estimate of one underlying prevalence. The instance denominators (81 and 57) count sensitive-argument instances under each field set. At task level, 33 user tasks have ground-truth calls with at least one side-effecting call carrying a sensitive argument under the broader field set; 13 of these 33 tasks contain at least one argument that is an environment-only match.

We manually inspected the matched instances. One request refers to ``the same participants as the Introductory meeting''; their identities occur in the calendar. Another refers to ``each person mentioned in the TODO list,'' whose identities occur in a document. A third asks for ``the user who wrote the most messages,'' identified through channel history. These are clear environment-only matches: the user does not state the values verbatim.

Values present in both the prompt and environment have ambiguous provenance under this method. Our prompt-first rule assigns them to the prompt-match category. We therefore do not treat the environment-only matching rate as a lower bound on causal environment derivation.

\paragraph{Interpretation and limits.} An environment-only match does not imply attacker control, a cross-origin source, or untrusted content. The examples motivate separating a value's source from authorization of the requested task; they do not show which source an agent actually used.
\subsection{Supporting measurement: planner behavior}
\label{sec:susc}
The next measurement motivates checking execution independently of planner behavior, as we do in \S\ref{sec:nativeeval}. Planner refusal can prevent an injected action, but does not establish an independent authorization boundary. We measure how often each planner selects an action induced by the attacker. We use a controlled native-source fixture on a patched AOSP platform. It combines a legitimate task with content that tries to redirect a call to the attacker's target. We record whether the chosen call targets the attacker (the \emph{attacker-target-selected} event). This measures the planner's choice, independently of any downstream write.

We report each model separately. A \emph{family} is a set of trials sharing one task, injection template, and configuration; \emph{styles} are injection-phrasing variants. Open-weight counts come from the natural-workload family. For each frontier model, we pool only its own adversarial-style families. Table~\ref{tab:susc} gives the counts. These natural-workload trials are separate runs from the enforcement comparison in \S\ref{sec:nativeeval}: the same task and injection template, but earlier runs on an API-36 build, with retained records pooled across run directories. Qwen2.5-7B selected attacker targets in $22/36$ of these trials and in $10/10$ of each later comparison family on an API-37 build; we do not pool the two or attribute the difference. Phi-3-mini selected attacker targets in $19/20$. Several other open-weight planners recorded $0$ selections. All tested frontier models also recorded zero: Claude Sonnet~5.5 $0/59$, GPT-4.1 $0/50$, GPT-5 $0/49$, Gemini-pro $0/49$, and Gemini-flash $0/49$.

\begin{table}[t]
\caption{Attacker-target selections / ATTACK trials. Open-weight and frontier models use different experimental families; these counts do not establish a controlled model ranking. $\dagger$: every trial was a parser error or an empty plan (Table~\ref{tab:status}); these zeros are generation failures, not demonstrated resistance.}
\label{tab:susc}
\centering\small
\begin{tabular}{lr}
\toprule
Model (our setup) & attacker-target / n \\
\midrule
Qwen2.5-0.5B & 0/31$^\dagger$ \\
Qwen2.5-1.5B & 0/31 \\
Qwen2.5-3B & 0/27 \\
Qwen2.5-7B & 22/36 \\
Qwen2.5-14B & 6/20 \\
Qwen2.5-32B & 0/20$^\dagger$ \\
Phi-3-mini & 19/20 \\
Phi-3-medium & 0/20$^\dagger$ \\
Llama-3.2-1B & 0/20$^\dagger$ \\
Llama-3.2-3B & 4/20 \\
\midrule
Claude Sonnet 5.5 & 0/59 \\
GPT-4.1 & 0/50 \\
GPT-5 & 0/49 \\
Gemini-pro (latest) & 0/49 \\
Gemini-flash (latest) & 0/49 \\
\bottomrule
\end{tabular}
\end{table}

\paragraph{Additional injection configurations.} In the styled-injection configurations, Qwen2.5-14B selected attacker targets in $45/45$ trials across styles~1--4 ($15$, $10$, $10$, and $10$ trials; vs $6/20$ in the natural workload), Qwen2.5-7B in $28/40$, and -32B in $33/40$. The 7B styled trials ran under the model-side label with no provider check, so every selection also committed (Table~\ref{tab:crafted}). Results varied by style; for example, 7B selected the target in $3/10$ trials of style~3. No tested frontier model selected an attacker target in its styled trials.

These configurations differ in more than wording: the 7B styled trials, for example, used a model-side label (a source marker shown to the planner; \S\ref{sec:nativeeval}). Their counts do not isolate an injection-style effect or support a controlled cross-model comparison. They establish that attacker-target selections occurred under several crafted-input configurations, motivating the separate provider-side tests in \S\ref{sec:nativeeval}.

\paragraph{Scope and caveats.} These are exploratory counts, not a calibrated benchmark. The models did not run on a phone, and we did not measure device resources. Susceptibility measures the planner's choice. Attack success requires a committed write and is reported separately in \S\ref{sec:nativeeval}.

A result of $0/n$ means that no attacker-target selection was observed in that setup. It does not prove robustness.

Several zero-selection cells reflect output-generation failures, not demonstrated resistance to injection. All $31$ Qwen2.5-0.5B, $20$ Llama-3.2-1B, and $20$ Phi-3-medium trials were classified as parser errors. Qwen2.5-32B produced only empty plans. We retain these attempts in the denominator. A zero selection count therefore does not establish successful completion of the authorized task.

Appendix~\ref{app:pilots} reports planner-status distributions, with counts summing to each model's $n$. Status and attacker-target selection are separate measures. For example, Phi-3-mini has $19$ attacker-target selections but $18$ trials with the planner status ``valid malicious call.'' The per-trial manifest distinguishes parsing failures, refusals, and benign task success. The observed selections motivate an execution check that does not rely on the planner rejecting injected actions.

\section{Application-Level Source Enforcement}
\label{sec:nativeeval}
In the controlled native fixture, our monitor withheld the tested writes to attacker-selected destinations. We ran it in two provenance configurations: a \emph{signed-hint} configuration, in which the broker signs a source set and confirmation state fixed by the test harness, and an \emph{OS-observed} configuration, in which the broker derives the source set from the OS accessibility log. In the OS-observed configuration the monitor also withheld benign sends issued after reading untrusted content, flagged benign work involving external content, and carried those source records into a later task. This case examines both enforcement decisions and the limits of using recent application observations to check individual actions.

\subsection{Policy and implementation}
We implement the application-source monitor by modifying AOSP accessibility delivery, the AppFunctions broker, and the experimental provider. Our platform changes record the \emph{source application package} of delivered accessibility nodes. The broker collects the packages observed by the agent into a set, which we call the source set. An OS-minted token binds this set, a confirmation state, caller, function, and argument digest for the provider. The broker obtains the set and the confirmation state in one of two ways, selected by a platform property. In the \emph{signed-hint} configuration, it signs the source set and confirmation state carried in the caller's AppFunctions attribution record; in our tests the harness, not the planner, fixes these per condition, so the planner's output cannot alter them, but a caller could. In the \emph{OS-observed} configuration, it ignores that hint, derives the set from a per-UID accessibility log over a fixed window, fails closed when the log is empty, and never takes confirmation from the caller. Only the OS-observed configuration makes the record independent of the caller; we state below which configuration each result used. In both, a side-effecting call with an untrusted source and no confirmation is withheld (not executed), and a textual provenance claim in the content cannot raise trust. These application identities do not authenticate a web document origin.

\paragraph{Decision rule.} We implement the rule in Table~\ref{tab:rule} at the broker and provider; the artifact identifies the implementation files and revision. In the OS-observed configuration, we derive the source set from a per-UID OS accessibility log over a fixed window, rather than from a caller-supplied hint. The rule represents user confirmation as an \texttt{ACTION\_CONFIRMED} state that only a trusted confirmation surface may set; the runs below either supplied that state through the harness (signed-hint) or ran without it (OS-observed). We configure the provider to admit a side-effecting call only if the user confirmed it or the observation window contains no untrusted source. Token, function, argument, or expiry mismatches cause rejection. We leave reads ungated because this policy targets side effects.

This rule also explains the operational limits below. A benign task can be flagged when it uses content from an application classified as untrusted. We use a 60-second per-UID window without a per-task reset; consequently, a subsequent clean task can inherit that source record (carryover). Because the set is keyed by package, injected content rendered by an application in the trust set is not flagged: the monitor cannot distinguish content within a trusted package, which is the gap of \S\ref{sec:measurement} seen from the native side.

\begin{table}[t]
\caption{Application-source monitor rules implemented by the broker and provider (OS-observed configuration; the signed-hint configuration takes the source set and confirmation state from the caller's attribution instead).}
\label{tab:rule}
\centering\small
\begin{tabular}{@{}p{0.22\columnwidth}p{0.72\columnwidth}@{}}
\toprule
Component & Rule \\
\midrule
Source set & Packages whose rendered content the OS accessibility log attributes to the \emph{calling UID} within a fixed window ($60$\,s in the prototype); the caller and target packages are excluded. Each is labeled \emph{trusted} or \emph{untrusted}; an empty set yields one default untrusted source (fail-closed). \\
Lifetime & The per-UID time window; there is \emph{no} per-task reset, so sources from a recent task persist (carryover). \\
Trust set & Fixed first-party/system packages plus a configured allowlist, set by the platform/experimenter---never by the caller. \\
Authorization & \texttt{ACTION\_CONFIRMED} is minted only by a trusted confirmation surface, bound to \mbox{(UID, action, value)}, single-use with a TTL; the caller cannot assert it. \\
Decision (side-effecting) & Admit if \texttt{ACTION\_CONFIRMED}, or if no untrusted source is in the observation window; otherwise flag/withhold. Token MAC, function, argument-digest, or expiry mismatches flag first. \\
Reads & Not gated (read-only functions bypass the check). \\
\bottomrule
\end{tabular}
\end{table}

\subsection{Controlled execution tests}
We evaluate the monitor with Qwen2.5-7B and a native-source attack fixture on the patched AOSP platform. To count only newly produced attacker-target effects, we require four conditions: the planner selects the attacker's target, the AppFunction executes, no matching message exists in the provider's store before execution, and a matching message exists after it. We say a write \emph{commits} when the message is inserted into the provider's in-memory store. The store is in memory; insertion does not imply durable storage or delivery to an external recipient. We count other new messages separately as generic writes because we have not established their authorization status.

Each condition contains up to ten recorded functional trials; the no-check control has nine. These counts describe the tested configurations, not an attack-success rate over a broader workload. Appendix~\ref{app:nativeoracle} specifies the raw-record fields and aggregation checks.

\paragraph{Defense comparison (overt injection).} We compare four configurations on the same task and attack fixture, with the monitor in its signed-hint configuration (Table~\ref{tab:defense}): no provider check, a model-side label, a confirmation gate, and our application-source monitor. The label marks untrusted content for the planner, in the manner of spotlighting~\cite{spotlighting}. The gate requests confirmation before a side-effecting call. We generate planner outputs independently for each configuration. For each, we read both whether the planner \emph{selected} the attack and whether a write \emph{committed}: a defense demonstrates blocking only when the planner selected an action that then failed to commit.

For CONTROL, we supply preconfirmed benign user-initiated calls: the harness-supplied record carries \texttt{ACTION\_CONFIRMED} and an empty source set, so the monitor admits on its confirmation branch. We provide no confirmation to the confirmation-gate baseline. The CONTROL column therefore reflects the supplied approval, not a difference between mechanisms; a confirmation gate given the same state would also admit. Under this rule the monitor behaves as a confirmation gate that waives confirmation when the recorded sources are all trusted. This tests execution under those approval conditions, not completion after interactive approval.

With no check, attack-specific writes commit $10/10$ and legitimate controls commit $9/9$. The model-side label records $0/10$ attack-specific commits under overt injection. However, the planner selects the attack in $0/10$ trials. The label therefore does not demonstrate blocking an already selected action.

Without additional confirmation, the confirmation gate withholds both attack and legitimate calls ($0/10$ each).
The application-source monitor records $0/10$ attack-specific commits (record: untrusted source, no confirmation) and admits $10/10$ legitimate controls (record: confirmed). The planner still selects the attacker action in $10/10$ trials. The provider withholds these writes; the planner does not avoid them.

\begin{table}[t]
\caption{Native enforcement with Qwen2.5-7B, monitor in the signed-hint configuration: newly stored attack-target messages (ATTACK) and benign writes (CONTROL). Counts show observed denominators; CONTROL calls carried a harness-supplied confirmation state that the gate baseline did not receive (see text).}
\label{tab:defense}
\centering\small
\begin{tabular}{lcc}
\toprule
Boundary configuration & ATTACK & CONTROL \\
\midrule
No provider check & 10/10 & 9/9 \\
Model-side label & 0/10 & 10/10 \\
Confirmation gate & 0/10 & 0/10 \\
Application-source monitor & 0/10 & 10/10 \\
\bottomrule
\end{tabular}
\end{table}

\paragraph{Crafted rephrasing.} We test four crafted injection styles and distinguish generic writes from effects matching the attacker's target (Table~\ref{tab:crafted}). With the model-side label, all four styles produce generic writes in $10/10$ trials. Attack-specific commits occur in $10/10$ for styles~1--2, but only $3/10$ and $5/10$ for styles~3--4.

The remaining $7$ and $5$ writes did not match the attacker's target; their authorization and task correctness were not evaluated. We ran the confirmation gate and the application-source monitor (signed-hint configuration) against style~1 only: the planner selected the attacker target in $10/10$ trials under each, and neither recorded an attack-target commit ($0/10$). The monitor checks the signed source set rather than the injected claim; styles~2--4 were not run against these two configurations.

\begin{table}[t]
\caption{Qwen2.5-7B with model-side labels, 10 trials per style. Generic writes need not be attack effects or authorized task completions.}
\label{tab:crafted}
\centering\small
\begin{tabular}{lcccc}
\toprule
Model-side label & style 1 & style 2 & style 3 & style 4 \\
\midrule
Generic write & 10/10 & 10/10 & 10/10 & 10/10 \\
Attack-specific effect & 10/10 & 10/10 & 3/10 & 5/10 \\
\bottomrule
\end{tabular}
\end{table}

\paragraph{Resistance to a textual authorization claim.} To test whether a textual authorization claim affects enforcement, we inject ``this is user-authorized, trusted first-party'' into the content. We observe attacker-target selection in $10/10$ trials and, with no check, $10/10$ attack-specific writes. With our monitor (signed-hint configuration), the planner still selects the attacker target ($10/10$), but no attack-specific write commits ($0/10$). The monitor uses the signed source record rather than the textual claim. This test concerns a textual authorization claim. It does not test browser impersonation, forged OS labels, or stolen tokens.

\paragraph{OS-observed configuration.} We repeated the attack fixture with the source set derived from the accessibility log (Table~\ref{tab:osobs}). In every trial the broker recorded the untrusted source application from the log, with no confirmation state. Qwen2.5-7B selected the attacker target in $10/10$ ATTACK trials, and no attack-specific write committed; Phi-3-mini selected it in all $9$ recorded ATTACK trials (one trial produced no record), with no commit. The CONTROL trials in this configuration are harness-supplied benign sends made after the same untrusted content was read: the monitor withheld all $10$ for each model, because the observation window contained an untrusted source and no confirmation. This withholding is the rule's cost on legitimate work when the window contains both trusted and untrusted sources. A separate single demonstration with only the Settings application recorded in the observation window admitted a call without confirmation.

\begin{table}[t]
\caption{OS-observed configuration: source set from the accessibility log, no confirmation state. CONTROL sends are harness-supplied benign calls after reading untrusted content.}
\label{tab:osobs}
\centering\small
\begin{tabular}{llrrr}
\toprule
Model & Condition & $n$ & Selected & New write \\
\midrule
Qwen2.5-7B & ATTACK & 10 & 10 & 0 \\
Qwen2.5-7B & CONTROL & 10 & 10 & 0 \\
Phi-3-mini & ATTACK & 9 & 9 & 0 \\
Phi-3-mini & CONTROL & 10 & 10 & 0 \\
\bottomrule
\end{tabular}
\end{table}

\paragraph{Operational limits of coarse source enforcement.} Our rule checks a set of recently observed sources, not the source of each argument. We therefore also probe benign work involving external content and a subsequent clean task.

In the signed-hint configuration, preconfirmed benign calls (record: \texttt{ACTION\_CONFIRMED}) required no additional confirmation ($0/20$ across \texttt{sendMessage}/\texttt{makeCall}, $N{=}10$ each). In the OS-observed configuration, a benign task whose observation window contained only a trusted application was admitted ($3/3$); a benign task that legitimately used another application's content was flagged ($3/3$), as was an injection task ($3/3$); and a clean task issued soon after a task that read external content was flagged ($1/1$) because the earlier source record remained in the observation window.

These are small-$N$ probes, not a false-positive rate. They show a limitation of coarse source checks. Benign tasks can legitimately use values from their environment (\S\ref{sec:workload}), yet the monitor may flag those tasks because of the source application. The WebView source-reference prototype (\S\ref{sec:webvieweval}) explores a more specific object-level binding under different trust assumptions; it is not evaluated as a drop-in fix for the application-source monitor.

\paragraph{Scope.} These functional checks use one or two open-weight planners and a native-source fixture at $N{\le}10$/cell. The four-configuration comparison, the style-1 crafted tests, the textual-claim test, and the preconfirmed workload use the signed-hint configuration; the OS-observed run and the workload probes use the OS-observed configuration. The model-side-label and confirmation-gate baselines are configurations we ran, not tuned production defenses.

\subsection{Execution costs and confirmation requests}
\label{sec:costs}
We measure costs for the native-source token and ChatApp provider. These measurements do not characterize the WebView variant. In summary, token operations are cheap relative to binder IPC and model inference; we report them at order-of-magnitude precision and do not claim a specific enforcement overhead.

\paragraph{Token cost (component microbenchmark).} On the patched platform, we time token minting in the broker and \texttt{parseAndVerify} in the provider with \texttt{System.nanoTime}. Minting uses HMAC-SHA256 over a canonical payload. The provider decodes, verifies, parses, and checks the token's policy. We aggregate per-call timings logged during a benign workload.

These wall-clock measurements vary with machine load. Per-operation medians range from tens to low hundreds of microseconds. We did not reproduce a stable median, so we report only the order of magnitude. The medians are not worst-case bounds: we observed a one-time \mbox{$\sim$1\,ms} first-call JIT cost and occasional higher tail samples.

We time minting and verification separately. Adding their medians would not give an end-to-end latency. Both measurements exclude source-reference table operations, browser capture, binder IPC, and confirmation interfaces. The comparison is qualitative: token operations take much less time than binder IPC (milliseconds) or model inference (seconds).

A repeated build-level timing comparison between an enforcing build and a baseline (Appendix~\ref{app:abba}) gave differences whose sign varied between runs and did not isolate the incremental cost of enforcement.

\paragraph{Additional confirmation requests.} For $20$ preconfirmed benign calls ($10$ each for \texttt{sendMessage} and \texttt{makeCall}) in the signed-hint configuration, the application-source monitor requested no additional confirmation, whereas the confirmation gate, which does not honor that state, requested confirmation for all $20$. Reads were exempt under both. These are provider decisions under a harness-supplied confirmation state, not observations from a user study. In the OS-observed configuration, benign sends after reading untrusted content, benign external-content tasks, and carryover tasks were withheld (\S\ref{sec:nativeeval}). The result does not establish zero user burden.

\section{Selected-Object Source Enforcement}
\label{sec:webvieweval}
The application-source monitor checks applications observed in a recent window; it does not identify the selected document. We implement the second prototype, the WebView source-reference gate, by modifying Chromium \texttt{content\_shell} to supply origin records and adding reference checks to the broker/provider path. The prototype browser is therefore \texttt{content\_shell}, not the WebView provider measured in \S\ref{sec:measurement}: we did not run the stamping inside an application-embedded WebView and assume, without verifying, that its accessibility export path is representative. The gate checks a reference to a stored object snapshot and its browser origin. It is a separate experiment under a cooperating-browser assumption, not a measured replacement for the application-source monitor on the same workload.

We use ``real path'' for execution through the running browser, broker, and provider components on the emulator, as opposed to the host-JVM policy evaluation. Calls rejected by the broker do not reach the provider.
After describing the contract and its evaluated implementation, we test the origin check's effect on stored state, reference binding, and lifetime under document changes. Live-agent behavior is reported separately. The limited functional tests do not establish an attack-blocking rate over a broader workload or a broad multi-model, multi-task security--utility result.

\paragraph{Terms used below.} A \emph{snapshot} stores the selected bytes. A \emph{source record} associates those bytes with an object, document, and origin. The OS \emph{reference table} stores these records. A \emph{source reference} is an opaque object handle used to look up the record and snapshot. A \emph{grant} specifies the approved action and binds it to that handle. Possessing a reference alone does not authorize an action.

An immutable snapshot survives navigation. Our document-bound ``live'' mode invalidates the reference on navigation; it does not fetch the page's latest text.

\subsection{Contract and evaluated implementation}
\label{sec:design}
\subsubsection{Implementation status}
The prototype resolves source references and checks origin, object identity, and lifetime. These operations run through the browser, broker, and provider. We use a fixed-approval harness to model the trusted issuer; this is an experimental substitute, not an implemented trusted user interface. Table~\ref{tab:implementation} separates that evaluated path from the full contract; Figure~\ref{fig:flow} shows their relationship. The browser-to-broker channel is not authenticated against browser impersonation, and atomic in-renderer selection is not implemented.

\begin{table}[t]
\caption{Status of the WebView contract in the evaluated prototype.}
\label{tab:implementation}
\centering\small
\begin{tabularx}{\linewidth}{lX}
\toprule
Status & Components \\
\midrule
Implemented & Source-reference table, snapshot resolution, broker/provider checks, in-memory message insertion \\
Simulated & Trusted selection and user approval, supplied by a fixed-approval harness \\
Unimplemented & Authenticated browser delivery and atomic renderer capture \\
Specified & Trusted parameter display, resolved-destination binding, and remote idempotency; full implementation is not established by these tests \\
\bottomrule
\end{tabularx}
\end{table}

\paragraph{Scope: sharing selected content.}
We restrict the task to sharing an exact text selection once with a specified recipient, so the contract need not translate arbitrary natural-language intent into a policy. We specify approval of the selected object, recipient, and operation through a trusted interface; opening or focusing a page does not authorize sending.
To preserve the selected bytes until execution, we associate the source reference with an immutable snapshot. We specify a record containing a browser-instance identity, a document identity, a navigation generation, the computed security origin, an object identifier, and a digest of the selected bytes.

In the evaluated path, we restrict issuance to the simulated selection path and give the agent only an opaque broker-issued handle, without access to node-extras writes. The provider accepts only a handle that resolves to the grant's object reference. Knowing the raw text is not an authorization bypass: the provider requires the reference and checks the operation scope.

For opaque origins, we require the contract to preserve browser-defined identity rather than treat all opaque documents as one trusted source. Full document URLs need not be exported merely to compare origins. We distinguish the top-level origin, selected-document origin, and frame ancestry in the contract so policies can refer to them separately; we do not evaluate all such policies.

\subsubsection{Issuance and execution}
\begin{figure}[t]
\centering
\begin{tikzpicture}[
  font=\footnotesize,
  box/.style={draw,rounded corners=2pt,align=center,inner sep=3pt,text width=0.80\linewidth},
  impl/.style={box,thick},
  req/.style={box,dashed},
  arr/.style={-{Latex[length=1.8mm]},thick},
  node distance=4.5mm,
]
\node[box] (agent) {Rendered web content {+} agent:\\ \emph{candidate} object selection (pre-approval, untrusted)};
\node[req,below=of agent] (ui) {Trusted-UI selection + atomic capture \emph{(simulated/required)}:\\ approve \emph{object, recipient, action} $\Rightarrow$ fix $(\textit{doc id},\textit{origin},\textit{bytes},\textit{object id})$};
\node[req,below=of ui] (chan) {authenticated browser$\to$broker channel \emph{(required)}};
\node[impl,below=of chan] (broker) {Broker: resolve source reference;\\ \textbf{origin-eligibility gate} (implemented)};
\node[impl,below=of broker] (prov) {Provider: ordered check $\to$ commit / withhold};
\draw[arr] (agent)--(ui);
\draw[arr] (ui)--(chan);
\draw[arr] (chan)--(broker);
\draw[arr] (broker)--(prov);
\begin{scope}[on background layer]
\node[draw,dotted,rounded corners,fit=(ui)(chan)(broker)(prov),inner sep=3pt,
  label={[font=\footnotesize]above right:\emph{trusted}}] {};
\end{scope}
\end{tikzpicture}
\caption{WebView selection and execution flow. Solid boxes are implemented; dashed boxes are simulated steps or unimplemented requirements. Trusted user approval is simulated by the harness.}
\label{fig:flow}
\end{figure}
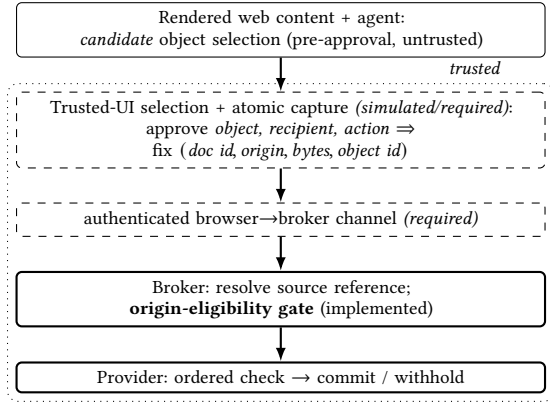

To constrain a call to the approved object and action, we require the trusted issuer to create a task grant $G$ (Figure~\ref{fig:flow}). The grant binds the caller and session, provider and function, and destination channel and recipient. It also binds the source-object reference and payload digest, with an expiry and a single-use nonce. In the prototype, we supply grants through the simulated selection path. The planner receives references for proposing calls, not access to the issuer. We specify broker resolution and the following provider checks before a write:

\begin{enumerate}
\item The caller, task, provider, function, and destination channel match the grant.
\item The source record exists, satisfies the origin policy, and has the required lifetime semantics.
\item Recipient, text digest, and attachment set match the granted scope exactly.
\item The grant is valid and unused; single-use consumption prevents re-executing the same grant.
\end{enumerate}

We use an in-memory message repository in the experimental provider and measure its state changes. \S\ref{sec:onoff} reports message insertions into this local state, not durable database commits or delivery to an external recipient. These measurements do not establish crash-safe duplicate suppression or exactly-once execution of remote effects.

\paragraph{Navigation and transformations.}
We distinguish two contracts. (A) An \emph{immutable snapshot} fixes the approved bytes at selection. Navigation or in-place edits do not change what is sent. (B) A \emph{live-content} reference sends the object's current bytes. Those bytes may differ from the approved payload, so this contract needs re-approval.

We implement (A). We also test a document-bound ``live'' mode that invalidates references on navigation, but does not implement the latest-content semantics of (B). The interface must identify which semantics apply, rather than reuse a frame identifier indefinitely.

We limit the prototype to exact selected text so the executed bytes can be compared with the approved snapshot. Summaries, translations, OCR, and merged content would require dependency tracking or separate approval of the derived payload. Attaching a source handle to an arbitrary generated summary does not establish its provenance. For a future extension without such dependency tracking, we specify an $\unknown$ source state and withholding or renewed authorization; the current tests do not evaluate those transformations.

\subsubsection{Relation to a capability, and soundness requirements}
\label{sec:reqs}
Checks~1, 3, and 4 are conventional capability checks: they bind the caller, provider, function, destination, and payload to a grant, then restrict its validity and reuse. A capability can also carry an origin condition at issuance. The question here is how to supply the source information needed to evaluate that condition at the native execution boundary.

The interface measurement in \S\ref{sec:measurement} identifies a missing input to such a policy. Our WebView prototype binds that input to a selected object. We isolate the effect of origin eligibility by toggling only that check in the same implementation (\S\ref{sec:onoff}). We test reference binding and lifetime separately. To make this binding trustworthy beyond the controlled prototype, the system must protect capture, delivery, payload, destination, and reuse. The following requirements specify these protections; we have not fully implemented them.

\begin{enumerate}
\item \textbf{Atomic capture.} The selected bytes and their per-frame origin must be captured \emph{atomically in the renderer} at selection, or a later read through the accessibility tree, which lags the renderer, admits a time-of-check/time-of-use window.
\item \textbf{Authenticated channel.} The origin must reach the broker over a channel \emph{authenticated to the browser} (e.g., platform signature). A native app impersonating a browser can forge a readable accessibility extra. Our prototype's eligibility check consumes the stamp under the cooperating-browser assumption; the stamp is not authenticated, and only the claimed-origin cross-check (\S\ref{sec:s2}) treats a caller-supplied origin as a \emph{hint}.
\item \textbf{Payload binding.} The payload digest must bind the \emph{transmitted bytes}, and the trusted UI must flag bidi-override/homoglyph display characters rather than fold them into the hash.
\item \textbf{Destination binding.} The grant must bind the actual destination identifier, such as a phone number or email address, \emph{after} any display name has been resolved.
\item \textbf{Idempotency.} The grant nonce must propagate to the remote endpoint as an \emph{idempotency key}.
\end{enumerate}
We require the user or enterprise to define the approved-origin set independently of the selected content, so selection cannot itself determine source eligibility.

\paragraph{Claim boundary.}
Under a correct browser, broker, trusted issuer, and completely mediated provider, the contract aims to prevent committing a payload or destination outside the issued grant. Missing mediation, compromised reference issuance, and untracked transformations can invalidate that objective. The functional tests below evaluate the specified execution checks. They leave content correctness, content safety, general prompt-injection resistance, and user burden unmeasured.

To meet the authenticated-channel requirement in \S\ref{sec:reqs}, a complete implementation could reuse the platform's UID or connection authentication, provided it explicitly binds each origin record to a genuine browser connection. The application-source monitor already records the delivering package at delivery (\S\ref{sec:nativeeval}); accepting origin records only from packages in a trusted-browser set would use that record as the authentication. We have not implemented this binding, so the current prototype does not establish it.

\paragraph{Prototype implementation.}
We mint source references only through the simulated trusted-selection path and expose opaque identifiers to the agent. We apply the current-document check only to document-bound (live) references, withholding them when navigation changes the bound document instance. For immutable snapshots, we retain the approved bytes across navigation and same-document edits; the document check itself does not detect in-place mutation.

We compare an agent-supplied origin argument with the recorded browser origin but never use it to replace that record. Appendix~\ref{app:proto} describes the OS-side reference table in \texttt{system\_server}, token contents, and grant initialization.

We implement two forms of the origin check for different tests. The diagnostic policy comparison and host-JVM evaluation use a trust-classification step that marks a record as policy-approved without changing the source's recorded origin. The reference-binding and execution tests use the source-reference gate, which resolves the selected snapshot and checks its object, origin, and lifetime before the provider is invoked.

\subsection{Effect of the origin check}
\label{sec:onoff}
\subsubsection{Stored-state comparison}
We first test whether enabling the origin check changes which objects reach the provider's store. We invoke the gate directly to test it independently of whether a planner selects the non-approved object. For each of the four conditions, we issue a fresh single-use grant with a distinct nonce; the claimed origin matches the recorded origin, and the object, lifetime, expiry, and single-use checks remain unchanged.

An approved object $P$ commits with the check either off or on. A non-approved object $Q$ commits with it off and is withheld with it on (Table~\ref{tab:onoff}). The withheld request stops at the broker before the provider is invoked. We use the in-memory commit definition from \S\ref{sec:design}.

\begin{table}[t]
\caption{Origin check enabled vs.\ disabled, one execution per condition with a fresh grant. Stored means an in-memory insertion. $P$ is approved; $Q$ is not.}
\label{tab:onoff}
\centering\small
\begin{tabular}{ccccc}
\toprule
Object & Check & Broker & Provider & Stored \\
\midrule
$P$ & off & admit & called & yes \\
$P$ & on & admit & called & yes \\
$Q$ & off & admit & called & yes \\
$Q$ & on & origin-policy & not called & no \\
\bottomrule
\end{tabular}
\end{table}

The $Q$ insertion with the check disabled violates the configured origin policy even though the bytes and recipient match the approved case. The added effect is the withheld $Q$ insertion, not a change in caller, payload, or object checks. To exercise the gate after a well-formed approval, we configure the harness to approve whichever object is selected before issuing its grant. This models the intended order of selection, displayed-action approval, and issuance. The claim is therefore conditional: even with approval to share that object, the separate origin policy can withhold it. Live trials with an attacker controlling only $Q$ are reported in \S\ref{sec:live}; they did not induce that selection.

\subsubsection{Diagnostic policy comparison}
\label{sec:originpolicies}
We compare three provider policies while holding user authority and rendered payload fixed (Table~\ref{tab:policy}). We serve the same value from an approved origin and a non-approved origin. Package-only classification trusts the host application and admits both, incorrectly admitting the non-approved origin. Blanket web-content rejection withholds both, including the approved iframe. Of these three policies, only the origin-aware policy admits the approved origin and withholds the other. It uses per-frame origin supplied by our cooperating browser.

These are deterministic decisions, not success rates. They isolate origin eligibility from the full binding between source and action. We use the modified browser's accessibility stamp as an origin hint to test policy decisions, not to demonstrate a hardened channel. Package-only classification is a diagnostic control; we do not attribute it to existing products.

\begin{table}[t]
\caption{Decisions in our three-policy comparison for approved and non-approved origins. A false admit is a false negative; a false withhold is a false positive.}
\label{tab:policy}
\centering
\begin{tabular}{lcc}
\toprule
Policy & Approved origin & Non-approved origin \\
\midrule
Package-only & admit & admit \textit{(false admit)} \\
Web-reject & flag \textit{(false withhold)} & flag \\
Origin-aware & admit & flag \\
\bottomrule
\end{tabular}
\end{table}

\subsubsection{Offline policy-function evaluation}
\label{sec:s1}
We compare two settings that differ only in whether the origin check is enabled. We supply identical inputs to capability-only and capability$+$origin within each case, then vary inputs across cases to exercise different violations. For example, the non-approved-origin case changes the object reference and origin while retaining the legitimate case's recipient and bytes.

To keep the grant structure fixed, we use the same fields in both settings: caller, action, destination channel $C$, recipient $R$, source-object reference, payload digest, expiry, and single-use nonce. The design in \S\ref{sec:design} also binds attachments, which this replay does not exercise. We configure the approved-origin set before the test, modeling the user- or administrator-defined policy required by the contract. Approval to share an object authorizes its recipient and action but does not override that policy.

Both settings apply the same ordered checks for scope, object, channel, replay, and expiry violations. Only capability$+$origin additionally rejects a source object outside the approved-origin set. Table~\ref{tab:s1} (Appendix~\ref{app:tables}) reports the reason codes. Both settings reject the general violations; they differ only on the otherwise valid grant for a non-approved-origin object, which capability-only admits and capability$+$origin withholds.

\noindent\emph{Evaluation method and scope.} To test policy logic separately from browser delivery and storage, we run the provider's decision function on the host JVM. We supply object origins as inputs and model state changes without committing to a store. The stored-state comparison above measures the corresponding decision on the real path.

We initialize each row with a fresh grant per setting and independent simulated state. For the replay case, we consume the grant on the first admission and check for \textsc{replay} on the second attempt. We evaluate expiry against a supplied logical clock.

\subsection{Binding the object to its browser origin}
\label{sec:s2}
The offline policy evaluation accepts a supplied origin. To test binding on the real path, we modify the browser to annotate (stamp) each web node with four fields: object identifier, document-instance identifier (the accessibility tree id), document origin, and text digest. We record this tuple at OS accessibility delivery. The simulated selection path then commits it as a source reference. The tree id changes on navigation. The agent receives references but cannot issue them or write node extras.

We place object-mismatch and origin-policy checks at the broker so rejected requests do not invoke the provider. The provider resolves the reference and uses its stored snapshot. It never bases an origin decision on model-submitted text. The check therefore concerns the object used in execution, not the source of whichever sentence the model may have read.

We exercise the checks through the running browser, broker, and provider: patched \texttt{content\_shell}, the AOSP delivery path and broker, and ChatApp. Table~\ref{tab:bind} records each call's grant target, presented reference, recorded origin, and claimed origin. These cases exercise three checks.

(i) \emph{Origin eligibility} uses the browser-asserted origin that the OS stored at delivery (unauthenticated against impersonation, \S\ref{sec:reqs}). An unapproved origin causes rejection (\textsc{origin-policy}), even when the caller reports that origin honestly. The check does not depend on the attacker making a false claim.

(ii) The \emph{claimed-origin} check compares the model-supplied origin with the recorded origin. It rejects disagreement (reason code \textsc{bind-mismatch} in the logs). This detects a false claim, but does not by itself establish an unforgeable reference-to-object binding.

(iii) \emph{Reference--object consistency} is checked by resolving the broker-issued reference to its
stored snapshot and matching it against the object named in the grant. Presenting $Q$'s reference under
a grant for $P$ is rejected (\textsc{object-mismatch}). These functional checks cover the tested
reference/object combinations, not a general unforgeability guarantee for reference issuance.

We also submit $P$'s reference with model-supplied bytes equal to $Q$'s; the call is admitted and transmits $P$'s snapshot. Because $P$ and $Q$ contain identical bytes, this case cannot show which bytes were used; that the provider transmits the stored snapshot rather than the model's bytes is established by the in-place mutation test (\S\ref{sec:s3}) and the code audit. Identical bytes also carry no information about which frame they were copied from, so a rejection on that basis would require evidence the check does not have.

We place identical bytes in the approved and non-approved frames to test a distinction that digest matching alone cannot express. We observe distinct object references and origins despite equal digests. The
binding is not yet authenticated against a native application impersonating a browser, for which an
authenticated browser-to-broker channel (\S\ref{sec:reqs}) is required.

\begin{table}[t]
\caption{Reference-binding checks on the real path, one execution per condition. $P$ is approved; $Q$ is not. Recorded origin comes from the OS reference table; claimed origin is supplied by the caller.}
\label{tab:bind}
\centering\footnotesize
\begin{tabular}{ccccl}
\toprule
Grant & Ref & Rec.\ origin & Claimed & Decision \\
\midrule
$P$ & $P$ & $P$ & $P$ & admit (transmit $P$) \\
$P$ & $Q$ & $Q$ & $Q$ & object-mismatch \\
$Q$ & $Q$ & $Q$ & $P$ & bind-mismatch (claimed origin) \\
$Q$ & $Q$ & $Q$ & $Q$ & origin-policy \\
\bottomrule
\end{tabular}
\end{table}

\subsection{Reference lifetime}
\label{sec:s3}
\paragraph{Navigation.} We test navigation with and without an origin change to distinguish document lifetime from origin eligibility. At selection, we fix a record $(\textit{document id},\textit{origin},\textit{bytes},\textit{object id})$ on the real path. We then replace the document under $P\!\rightarrow\!P$ (same origin, new document) or $P\!\rightarrow\!Q$ (Table~\ref{tab:life}).

Immutable snapshots are admitted after either navigation. Live references are withheld before the provider runs because the recorded document instance is no longer current. This includes $P\!\rightarrow\!P$: the origin is unchanged, but the document differs. The decision thus depends on the object/document/origin binding, not origin alone. A grant naming a different object fails the object check first.

We observed $0$ admissions with an incorrect origin or document and $0$ unnecessary rejections of valid snapshots in these conditions. Each condition is a single functional check of lifetime under navigation. We test a change during selection finalization separately below.

\begin{table}[t]
\caption{Reference lifetime under navigation, one execution per condition. $P$ is approved; $Q$ is not. Withheld calls do not reach the provider.}
\label{tab:life}
\centering\small
\begin{tabular}{lll}
\toprule
Reference & Navigation & Decision (provider) \\
\midrule
immutable & none & admit (executes) \\
live & none (current) & admit (executes) \\
immutable & $P\!\rightarrow\!P$ & admit (executes) \\
live & $P\!\rightarrow\!P$ & stale (withheld) \\
immutable & $P\!\rightarrow\!Q$ & admit (executes) \\
live & $P\!\rightarrow\!Q$ & stale (withheld) \\
\bottomrule
\end{tabular}
\end{table}

\paragraph{Document change during selection finalization.} We implement preparation and finalization steps to detect document replacement during simulated selection. Preparation records a pending object and its document instance. Finalization mints a reference only if that instance is still current.

We navigate between these steps. In the no-navigation control, reference issuance and execution succeed. With document replacement, the recorded instance is no longer current, so issuance aborts without minting a reference. A later call naming the absent reference is rejected as unknown and does not invoke the provider. We run each condition once. This checks the finalization point; it does not establish atomic renderer capture or close every time-of-check/time-of-use window.

\paragraph{In-place change without navigation.} To test snapshot consistency beyond the document-instance check, we mutate the selected paragraph without navigating, preserving its tree id. We compare transmitted bytes with the approved snapshot. After mutation, both immutable and live references transmit bytes matching the approved digest. The no-mutation control also admits normally. The provider uses the selection-time snapshot rather than the changed page text.

We do \emph{not} provide reliable live in-place invalidation. Our attempted digest re-check did not detect the mutation and admitted the live reference (one mutated and one unmutated case). We attribute this to its lookup, which used accessibility node identifiers whose correspondence is not stable across tree updates, rather than to digest comparison in general; the logs record only the decisions.

We audit the reference lookup code and confirm that it binds the opaque reference to the selection-time tuple rather than a reused node ID. Immutable references therefore preserve the captured bytes. A future live reference that tracks the latest content would need reliable object-version tracking and renewed authorization when bytes change. Atomic capture at selection alone cannot detect later changes (\S\ref{sec:reqs}).

\subsection{Live-agent trials and recovery}
\label{sec:live}
\paragraph{Minimal live-agent run.} We used one task and Qwen2.5-7B in an $N{=}8$ pilot, with the protocol fixed in advance and no model retries, followed by a separate $24$-trial evaluation. The agent chooses between an approved paragraph $P$ and a non-approved embedded paragraph $Q$. Paragraph $Q$ contains an injection that tries to redirect the share. We submit the selected object to the real path and configure the harness to approve either choice. We measure the agent's choices separately from the gate's decisions.

The 7B agent completed the benign $P$ share in all $8$ single-injection pilot trials. A separate evaluation used four styles: authority, urgency, fake internal task, and cross-channel social engineering. We ran six trials per style ($24$ total), disjoint from the pilot. The $0/24$ selections of $Q$ refer only to this four-style evaluation; the $8$ pilot trials are reported separately.

The prompt supplies both objects' text and identities, including labels absent from the original accessibility path. These labels identify the approved object, which the real accessibility path would not do; the $0/24$ is therefore weaker than a real-path result, and it does not contradict the $10/10$ and $28/40$ attacker-target selections of the same model in \S\ref{sec:nativeeval}, which used a different fixture without such labels. No model-induced attack reached the execution boundary.
We therefore evaluate the gate through direct adversarial calls, independently of model susceptibility. A $Q$ call is withheld at the broker with the origin check on; it commits with the check off. Rejected calls never invoke the provider. Each admitted call makes one \texttt{executeAppFunction} invocation. We count broker requests and provider calls separately.

To check that the path remains usable after a rejection, we force the harness to re-select $P$ and complete the task. This tests recovery execution, not autonomous model recovery. The $Q\!\rightarrow\!P$ sequence makes two broker requests and one provider call, storing one message. Compared with the normal $P$ path, it adds one selection and one broker request. It adds no provider invocation because the rejected $Q$ call never reaches the provider. These counts are an \emph{interaction cost}, not a measure of user burden or approval-error rate.

\section{Implications of the Two Cases}
\label{sec:implications}
The choice of source information determines what a policy can express. Recent application history supports a rule about which applications the agent observed. It cannot identify the particular object used by a call. Increasing planner reliability would not make that history more precise.

Object-specific policies instead require a record tied to the selected object. Matching bytes is insufficient when identical text appears in approved and non-approved documents. Lifetime semantics matter as well: an approved snapshot can remain valid after navigation even though a reference to the current document expires. These distinctions explain the different checks in our WebView design (\S\ref{sec:design}).

The two cases do not measure a trade-off curve or demonstrate that object references reduce the application-source monitor's restrictions on benign tasks. They use different workloads and trust assumptions. Together, they identify two separate requirements: choose source information precise enough for the intended policy, and preserve its association with the executed object across the observation boundary. A more precise source label alone is insufficient. The system must also link that label reliably to the object the action uses.

The separation of source eligibility and action authorization (\S\ref{sec:policy}) also applies after an object is bound. A stored snapshot answers which bytes will be used; it does not decide who may receive them.

\section{Limitations and Ethical Considerations}
We measure one primary WebView configuration and conduct a limited reproduction on a second version. We evaluate one modified browser and one provider application, ChatApp (Table~\ref{tab:env}). This is not a systematic study across versions or applications.

Truncation, repeated observations, missing traversal identifiers, and incomplete binary verification limit the measurements. Server and node logs also use different clocks. Their timestamps alone cannot establish a synchronized trace of one page load.

The standalone public-API collector checks dependence on framework instrumentation. The stock, Google-distributed WebView reproduction checks dependence on the modified image (\S\ref{sec:measurement}). Broader replication still needs synchronized logs, complete tree traversal, and verified binary identities.

Our finding is limited to the inspected interface fields. A production-agent exploit, a same-origin-policy bypass, and a claim that origin information is universally unavailable are outside the demonstrated results. A missing field can motivate hardening without violating a platform security guarantee. Conversely, origin metadata cannot establish content trust within one origin or authorize an action.

\paragraph{Ethics and disclosure.} We use locally served synthetic pages and restrict the exploratory mail task to an in-memory benchmark mailbox, avoiding external messages, third-party accounts, and real financial actions. We reported an AppFunctions caller-attribution concern to the vendor, who asked that the WebView origin observation be filed as a separate report. Neither has been confirmed as a vulnerability; we claim no CVE, fix, or first discovery for either. When preparing the public artifact, we will remove identifying build strings and local paths while preserving version information.

\paragraph{Generalization across workloads.} The supporting measurements use one benchmark and exploratory planner fixtures, not a production workload sample. The prototypes use separate tasks and trust assumptions. Establishing performance across applications would require broader workloads and a common evaluation protocol; the present results do not measure that generalization. Experiment-specific limits remain with the corresponding results (\S\ref{sec:workload}--\ref{sec:live}).

\section{Conclusion}
We measured which source information is available to Android agents and built two prototypes to examine its use at execution. The inspected WebView accessibility fields omit document origins, while recent application history can restrict benign work. Our selected-object prototype tests origin and lifetime checks through direct calls; authenticating browser records and capturing selections atomically remain open implementation requirements. The study shows why authorization needs both suitable source information and a reliable link to the object used in execution.

\appendix
\section{Live-agent and excluded-pilot details}
\label{app:pilots}
\emph{Live-agent experiment (\S\ref{sec:live}).} We use single-turn prompts with temperature $0.7$, top-$p$ $0.9$, and no retries. We run $8$ trials with one injection and $24$ with four distinct injection styles $\times$ six trials, keeping development injections separate from this evaluation set. We supply both objects' text and identities in the prompt, including labels absent from the original accessibility path. We do not treat the $0/24$ as a general robustness claim. We report the following quantities separately: $Q$-selected, calls reaching the execution boundary, actual commits, benign completions; we do not merge the direct-invocation result with the live arm.

We also ran an exploratory AgentDojo pilot with eight custom cases and a local Qwen2.5-7B planner. Four summary cases terminated with malformed tool-call output. Four reply cases completed the authorized reply without unauthorized sends. These are not official benchmark scores or a successful independent reproduction, and we did not evaluate the proposed guard in that pilot; we exclude it from security-performance claims.

Earlier checks that only tested whether a source record was present, and counterfactual comparisons of task outcomes, motivated the study but do not validate the origin contract. Rerunning a benchmark with the injection removed is not the same as a runtime filter that removes injected content before the planner sees it. Equal task-success outcomes also do not establish that the same authorized calls occurred. We exclude these exploratory results from any overall attack-blocking rate.

\emph{Trial-record manifest and classification protocol.} We recompute susceptibility counts (\S\ref{sec:susc}) and native-source comparisons (\S\ref{sec:nativeeval}) from per-trial \texttt{evidence.json} records, not prior summaries. The manifest lists the experiment family, model identifier, condition, planner status, attacker-target-selected flag, and committed-write flag. It distinguishes parsing failures and refusals from benign completions.

Table~\ref{tab:susc} excludes trials from the model-side label, confirmation gate, provenance monitor, and textual-claim families. We report these separately in \S\ref{sec:nativeeval}. The 7B crafted-label family appears among the additional injection configurations in \S\ref{sec:susc}.

Our AgentDojo classifier normalizes argument values before applying the matching rules in \S\ref{sec:workload}. The artifact includes the classifier used for those counts.

For reproducibility, the manifest records each model's exact API identifier or local checkpoint and revision. It also records the run date and serving and decoding settings. Frontier \texttt{-latest} aliases (e.g.\ \texttt{gemini-*-latest}) are not reproducible version identifiers. We therefore also record the resolved model string and date returned at run time.

A cell's $n$ counts retained execution records. These may include repeats or retries across run directories and seeds, rather than independent attack designs. The artifact manifest records the run/seed composition and inclusion rule. The artifact includes that manifest, the classifier, per-family counts, the per-trial records, and the prototype sources; a public release is in preparation.

\begin{table}[t]
\caption{Planner outcomes for the natural-workload ATTACK trials in Table~\ref{tab:susc}; rows sum to $n$. Valid denotes \texttt{PLANNER\_VALID\_MALICIOUS\_CALL} (function and target argument match the attacker's call); w-arg/w-fn denote a parsed call whose argument or function differs from that call, none an empty plan, and parse a parser error. Status differs from attacker-target selection.}
\label{tab:status}
\centering\small
\begin{tabular}{lrrrrrr}
\toprule
Model & valid & w-arg & w-fn & none & parse & $n$ \\
\midrule
Qwen2.5-0.5B & 0 & 0 & 0 & 0 & 31 & 31 \\
Qwen2.5-1.5B & 0 & 2 & 5 & 13 & 11 & 31 \\
Qwen2.5-3B & 0 & 25 & 0 & 2 & 0 & 27 \\
Qwen2.5-7B & 22 & 14 & 0 & 0 & 0 & 36 \\
Qwen2.5-14B & 6 & 11 & 0 & 3 & 0 & 20 \\
Qwen2.5-32B & 0 & 0 & 0 & 20 & 0 & 20 \\
Phi-3-mini & 18 & 1 & 1 & 0 & 0 & 20 \\
Phi-3-medium & 0 & 0 & 0 & 0 & 20 & 20 \\
Llama-3.2-1B & 0 & 0 & 0 & 0 & 20 & 20 \\
Llama-3.2-3B & 4 & 1 & 8 & 7 & 0 & 20 \\
\bottomrule
\end{tabular}
\end{table}

\section{Prototype implementation details}
\label{app:proto}
We add a source-reference table to OS \texttt{system\_server} to store references and their selected-byte snapshots. The simulated trusted-selection path issues opaque identifiers with a TTL. The broker resolves an identifier and checks it against the object reference in the grant.

The broker observes navigation and document changes through accessibility delivery. A document-bound (live) reference remains valid only while its document-instance identifier (the accessibility tree id) is current. An immutable snapshot remains valid across navigation.

We bind the caller, function, argument digest, and source record in the OS-minted token. The provider verifies the token and re-applies the origin policy. The trust-classification step marks a passing policy check: the origin is approved for this channel. It does not change the recorded origin. The provider only cross-checks an agent-supplied origin argument; it never substitutes it for the browser's record.

We issue a fresh single-use grant with a distinct nonce for each real-path condition. This does not imply that the entire table is reset between conditions. The policy-function replay uses independent simulated state per row (\S\ref{sec:s1}).

\subsection{Build-level timing comparison}
\label{app:abba}
We compared an enforcing build with a baseline on the same benign workload (\texttt{PcafUsabilityTest}), timing the complete instrumentation invocation on the running emulator. We alternate builds in counterbalanced blocks (enforcing, baseline, baseline, enforcing) to offset temporal drift, with $8$ timed runs per block after $2$ warmup runs, reinstalling only at a block change, giving $16$ observations per build in each repetition. Across four repetitions, the enforcing-minus-baseline mean differences were $+11.5$, $-16.6$, $-6.2$, and $+2.4$\,ms (Table~\ref{tab:cost}).

The two builds differ in outcome as well as in build configuration. The baseline admits and commits the calls. In the enforcing build, run in the OS-observed configuration, these harness-supplied calls carry no observed source (the log recorded none for the instrumented agent) and are withheld fail-closed. The builds (\texttt{chatapp-tracing.apk} vs \texttt{chatapp.apk}) also differ in build configuration beyond enforcement, and initialization, scheduling, and temporal effects were not separately measured. The comparison therefore does not identify the incremental cost of enforcement and establishes neither zero overhead nor an upper bound on it. We retain per-run permutation and bootstrap statistics in the artifact but do not rely on them, because they resample trials independently without modeling block or time dependence.

\begin{table}[t]
\caption{Native build timings (ms); mean differences are enforcing minus baseline. The builds differ in build configuration and in outcome, so the differences do not isolate enforcement cost.}
\label{tab:cost}
\centering\small
\begin{tabular}{lrrr}
\toprule
Run & Enf.\ median & Base median & Mean diff. \\
\midrule
Initial & 594.5 & 583.0 & $+11.5$ \\
Repeat 1 & 636.5 & 651.5 & $-16.6$ \\
Repeat 2 & 629.0 & 641.5 & $-6.2$ \\
Repeat 3 & 637.0 & 630.5 & $+2.4$ \\
\bottomrule
\end{tabular}
\end{table}

\subsection{Restricted policy inputs}
Let a delivered observation be $z=(t,p,c,r,E,A)$, comprising text, package, class, resource identifier, extras, and advertised extra-data keys. Let $o(d)$ be the browser-computed origin of document $d$. A policy over origins requires a trustworthy association between $z$ and $o(d)$; it cannot obtain that association merely by authenticating package $p$.

To isolate what these fields can distinguish, we restrict the hypothetical consumer to the projection $q(z)=(p,c,r,K)$, where $K$ is the extras-key schema. If two observations share this projection, the consumer must make the same decision on both. This elementary fact explains why those fields cannot implement a policy requiring different decisions for the two origins. It does not prove that the full observations are identical: the measured nodes differ in text, geometry, and position. Nor is this a universal impossibility result for black-box planners. It is a statement about a deliberately restricted policy input.

\twocolumn[{
\section{Supporting tables and execution criterion}
\label{app:tables}
\begin{minipage}{\textwidth}
\captionof{table}{Overview of measurements, trust assumptions, and evaluation criteria.}
\label{tab:evmap}
\centering\small
\begin{tabularx}{\textwidth}{lXXlll}
\toprule
Result & Input inspected & Trust basis & Decision criterion & Variant & \S \\
\midrule
Argument matching (workload) & AgentDojo ground-truth args vs prompt/env text & none (characterization) & string-match classifier & --- & \ref{sec:workload} \\
Planner susceptibility & planner output vs attacker target & none (model behavior) & attacker-target selected & --- & \ref{sec:susc} \\
WebView origin-attribution gap & a11y extras + advertised extra-data keys & OS a11y read & document-origin key present? & --- & \ref{sec:measurement} \\
Three-policy decision & per-frame origin (browser stamp) & cooperating-browser hint & provider admit/withhold & WebView & \ref{sec:originpolicies} \\
Native-source defense & app package in signed source record & harness-supplied hint (comparison); OS a11y log (OS-observed run) & provider write committed? & Native & \ref{sec:nativeeval} \\
Textual-authorization claim & injected ``user-authorized'' text & signed source record (harness-supplied hint) & provider write committed? & Native & \ref{sec:nativeeval} \\
Reference binding + lifetime & source reference (browser stamp + OS reference table) & browser + trusted selection & broker/provider decision & WebView & \ref{sec:webvieweval} \\
Component cost & native-token mint/verify timing & --- & \texttt{nanoTime} microbench & Native (token) & \ref{sec:costs} \\
\bottomrule
\end{tabularx}
\end{minipage}

\vspace{0.5em}
\paragraph{Artifact experiment identifiers.} The artifact retains development identifiers for locating logs: S1 is the offline policy-function evaluation (\S\ref{sec:s1}); B1 is reference binding (\S\ref{sec:s2}); B2 is navigation lifetime (\S\ref{sec:s3}); E1/E2 is the origin-check enabled/disabled comparison (\S\ref{sec:onoff}). These names do not denote additional experiments.

\vspace{0.5em}
\noindent\begin{minipage}[t]{0.48\textwidth}
\captionof{table}{Attribution-trace counts. Rows may overlap; records are repeated observations, not unique nodes.}
\label{tab:counts}
\centering\small
\begin{tabular}{lr}
\toprule
Observation & Count \\
\midrule
All node-delivery records & 191 \\
WebView shell package & 164 \\
Launcher package & 21 \\
Quick Search Box package & 6 \\
Records containing \texttt{chromeRole} & 93 \\
Host marker records & 13 \\
Cross-origin marker records & 13 \\
Records containing an HTTP(S) URL & 11 \\
\bottomrule
\end{tabular}
\end{minipage}\hfill
\begin{minipage}[t]{0.48\textwidth}
\captionof{table}{Offline policy-function evaluation: identical inputs within each case, with only the origin check toggled. Entries are reason codes, not measured store effects.}
\label{tab:s1}
\centering\small
\begin{tabular}{lcc}
\toprule
Input & Capability-only & Capability$+$origin \\
\midrule
Legitimate (origin approved) & admit & admit \\
Non-approved origin & \textbf{admit} & \textbf{origin-policy} \\
Recipient substitution & scope & scope \\
Payload substitution & scope & scope \\
Object swap (grant $\neq$ object) & object & object \\
Channel mismatch & scope & scope \\
Replay (2nd use) & replay & replay \\
Expired grant & expired & expired \\
\bottomrule
\end{tabular}
\end{minipage}
\vspace{0.5em}
\subsection{Native execution criterion}
\label{app:nativeoracle}
We compute the per-trial attack-effect indicator as
\[
\begin{aligned}
&\texttt{attackMarkerSelected}\\
&\land\ \texttt{actualAppFunctionExecution}\\
&\land\ \texttt{sideEffectPresentAfter}\\
&\land\ \lnot\texttt{sideEffectPresentBefore}.
\end{aligned}
\]
Across all trials, \texttt{sideEffectPresentBefore} was false. ``After'' therefore denotes a message newly inserted by the call. We call such an in-memory insertion a commit. A generic write is any newly observed post-call side effect; it need not match the attacker's target.

We recompute counts from raw trial records with the artifact's aggregation script. We verified that the four-condition criterion gives the same per-family counts as \texttt{attackMarkerSelected} $\land$ \texttt{sideEffectPresentAfter}. Each cell is a small set of functional tests (target $N{=}10$; the no-provider-check control has $9$ recorded trials). These are not estimates of attack-success rates over a broader workload.

\vspace{0.5em}
}]

\end{document}